\documentclass[twocolumn,floatfix,superscriptaddress,a4paper,showpacs,showkeys,nofootinbib,reprint,prc,noeprint]{revtex4-1}
\usepackage{epsfig}
\usepackage{latexsym}
\usepackage[colorlinks=true,linktocpage=true,linkcolor=blue,citecolor=blue,allcolors=blue]{hyperref}
\usepackage{url}
\usepackage[utf8]{inputenc}
\usepackage{enumerate}
\usepackage{color}
\usepackage{xcolor}

\usepackage{amsmath}
\usepackage{amssymb}

\usepackage[english]{babel}
\usepackage{hyperref}
\usepackage{url}
\usepackage{needspace}
\hypersetup{
   colorlinks=true, 
  linktoc=all,     
  linkcolor=blue,  
}

\begin{document}

 \title{
Energy dependence of light hypernuclei production in heavy-ion collisions\\ from a coalescence and statistical-thermal model perspective
 }

\author{Tom Reichert}
\affiliation{Institut f\"{u}r Theoretische Physik, Goethe Universit\"{a}t Frankfurt, Max-von-Laue-Str. 1, D-60438 Frankfurt am Main, Germany}
\affiliation{
Helmholtz Research Academy Hesse for FAIR (HFHF), GSI Helmholtzzentrum f\"ur Schwerionenforschung GmbH, Campus Frankfurt, Max-von-Laue-Str. 12, 60438 Frankfurt am Main, Germany}

\author{Jan Steinheimer}
\affiliation{Frankfurt Institute for Advanced Studies, Ruth-Moufang-Str. 1, D-60438 Frankfurt am Main, Germany}

\author{Volodymyr Vovchenko}
\affiliation{Institute for Nuclear Theory, University of Washington, Seattle, WA 98195-1550, USA}
\affiliation{Frankfurt Institute for Advanced Studies, Ruth-Moufang-Str. 1, D-60438 Frankfurt am Main, Germany}

\author{Benjamin D\"onigus}
\affiliation{Institut f\"{u}r Kernphysik, Goethe Universit\"{a}t Frankfurt, Max-von-Laue-Str. 1, D-60438 Frankfurt am Main, Germany}

\author{Marcus Bleicher}
\affiliation{Institut f\"{u}r Theoretische Physik, Goethe Universit\"{a}t Frankfurt, Max-von-Laue-Str. 1, D-60438 Frankfurt am Main, Germany}
\affiliation{
Helmholtz Research Academy Hesse for FAIR (HFHF), GSI Helmholtzzentrum f\"ur Schwerionenforschung GmbH, Campus Frankfurt, Max-von-Laue-Str. 12, 60438 Frankfurt am Main, Germany}

\date{\today}

\begin{abstract}
A comparison of light hypernuclei production, from UrQMD+coalescence and the thermal model, in heavy ion collisions over a wide range of beam energies and system sizes is presented. We find that both approaches provide generally similar results, with differences in specific details. Especially the ratios of hypertriton to $\Lambda$ are affected by both the source radius $\Delta r$ of the coalescence procedure as well as canonical effects. 
On the other hand, the double ratio $S_3$ is almost independent of canonical effects, which is in contrast to coalescence. Thus, both the beam energy dependence and centrality dependence of $S_3$ can be used to constrain the hypertriton source radius. To do so the currently available data is not yet sufficient. Elliptic flow is shown to be unaffected by the source size of the nuclei and an almost perfect mass scaling of the elliptic flow is observed. Our predictions further suggest that the existence of the H-dibaryon ($\Lambda\Lambda$) seems ruled out by ALICE data. 

\end{abstract}

\maketitle

\section{Introduction}

Hypernuclei, ordinary nuclei with at least one bound hyperon, are an important topic of nuclear physics \cite{Greiner:2001sy}. Understanding the creation and properties of hypernuclei can help in the understanding of the strong interaction and the role of flavor symmetry, relevant for nuclear structure but also the nuclear equation of state at high density. Heavy-ion reactions at relativistic energies are an abundant source of strangeness and therefore well suited for the production of light hypernuclei. 

Recently, several heavy-ion experiments have published data on the production of (anti-)hypernuclei and on their properties, e.g. the lifetime~\cite{STAR:2010gyg,Rappold:2013fic,Rappold:2014jqa,ALICE:2015oer,STAR:2017gxa,STAR:2019wjm,ALICE:2019vlx,STAR:2021orx}.

The lifetime measured in these experiments was found to be significantly below the free $\Lambda$ lifetime which was not expected from Faddeev-type calculations~\cite{Kamada:1997rv,Gloeckle:1998ty}. This lead to the so-called hypertriton puzzle, i.e. a significant deviation of the hypertriton (${}^{3}_\Lambda$H) lifetime from the lifetime of the free $\Lambda$. Currently, the tension between this expectation and the data is about 4.2$\sigma$~\cite{Donigus:2020fon,doenigusSQM2022}. Nevertheless, the measured properties of the hypertriton lead to consequences also for its production. 

In particular, it was suggested that the specific structure of the hypertriton, i.e. a small deuteron core with a weakly bound $\Lambda$ ($B_{\Lambda} = 0.162 \pm 0.044$ MeV~\cite{HypernuclearDataBase}, $B_\Lambda$ being the so-called $\Lambda$ separation energy), would lead to observable consequences in the system size dependence of hypertriton production~\cite{Sun:2018mqq}.  

Since measurements of hypernuclei are yet scarce and done at widely varying beam energies, it can be useful to investigate their production properties in models that can span such a big range of energies and system sizes in a consistent manner. Then systematic trends in the dependence of the hypernuclei production on its properties can be extracted. 
In this work, we will attempt just that and present calculations of light single and doubly-strange hypernuclei in heavy-ion collisions over a broad range of beam energies and system sizes. For a complete picture we will compare production rates from a (canonical) thermal model (Thermal-FIST) with those obtained from a coalescence model based on freeze-out distributions modeled with the UrQMD (hybrid-)model. Finally, we identify an observable which shows the most promising dependence on the hypernuclei production properties.

\section{Methods}

To make realistic predictions for the production rates and properties of hypernuclei, dynamic simulations are necessary. Since we want to study a broad range of beam energies and system sizes we will employ the UrQMD transport model to simulate the underlying hadron phase space distributions. The UrQMD model is a microscopic transport model based on the propagation and 2-body scattering of hadrons according to a geometrical interpretation of the scattering cross sections \cite{Bass:1998ca,Bleicher:1999xi}. For beam energies of $\sqrt{s_{\mathrm{NN}}}< 10$ GeV, this model provides a good description of experimental observables and measured hadron spectra in heavy ion collisions. For higher beam energies the model significantly underestimates the flow created \cite{Petersen:2007ca} as well as the strangeness produced \cite{Steinheimer:2011mp} which is why a so-called hybrid-model was established in which the dense phase is described by an ideal fluid dynamical simulation \cite{Petersen:2008dd}. 

In the hybrid description, the transition from the fluid description back to the transport description occurs on an iso-energy-density hypersurface  $\epsilon = 3 \epsilon_0$, where $\epsilon_0  \approx 145$~MeV/fm$^3$. The hypersurface is then used to sample hadrons according to the Cooper-Frye equation \cite{Cooper:1974mv,Huovinen:2012is} which then continue to interact within the cascade part of the UrQMD model, until reactions cease and kinetic freeze-out is reached. For the hydro part we use an equation of state that contains a smooth crossover between a hadron resonance gas and a deconfined quark-gluon-plasma~\cite{Motornenko:2019arp}.

\subsection{Thermal model}

The thermal model of particle production in heavy-ion collisions assumes that their primordial abundances are fixed at the stage of chemical freeze-out and correspond to the hadron resonance gas model in chemical equilibrium~\cite{Cleymans:2005xv,Andronic:2018qqt}. 
The only changes to the final abundances come from decay feed-down.
The model parameters -- the temperature $T$, baryochemical potential $\mu_B$, and the freeze-out volume $V$ -- are extracted at each collision energy by fitting the experimental data.
The thermal model is used to describe light (anti-)(hyper-)nuclei production by incorporating these objects as explicit degrees of freedom in the partition function~\cite{Andronic:2010qu,Donigus:2020ctf}.
Under the assumption that chemical freeze-out of light nuclei happens simultaneously with other hadrons\footnote{This assumption can be relaxed to allow light nuclei production at later stages, if one takes partial chemical equilibrium into account. In such a scenario one obtains similar results as in the standard thermal model~\cite{Vovchenko:2019aoz,Neidig:2021bal}.}, the model provides  predictions for light nuclei abundances in central collisions of heavy ions without introducing further parameters. 
In many cases, the model shows good agreement with the experiment~\cite{ALICE:2015wav,NA49:2016qvu}.
Augmented with the canonical treatment of baryon number conservation, the model can also describe features of light nuclei production in small systems at the LHC~\cite{Vovchenko:2018fiy}.

In the present work, we confront the predictions of the UrQMD coalescence approach both with the thermal model and experimental data.
For making predictions of the midrapidity yields $dN/dy$ at various collision energies, one has to specify the thermal model parameters $T$, $\mu_B$, and $V$ as a function of $\sqrt{s_{\mathrm{NN}}}$.
To this end, we utilize the chemical freeze-out curve of Ref.~\cite{Vovchenko:2015idt} which parameterizes the collision energy dependence of the temperature and baryochemical potential, $T(\sqrt{s_{\mathrm{NN}}})$ and $\mu_B(\sqrt{s_{\mathrm{NN}}})$.
In principle, this parametrization is sufficient to study the collision energy dependence of any yield ratio since the remaining volume parameter $V(\sqrt{s_{\mathrm{NN}}})$ cancels out in any such ratio.
Nevertheless, it can also be helpful to study thermal model predictions for absolute yields, for which one has to additionally specify the $V(\sqrt{s_{\mathrm{NN}}})$ dependence.
We fix $V(\sqrt{s_{\mathrm{NN}}})$ for 0-5\% central Au-Au/Pb-Pb collisions in the following way.
First, we use the world data~\cite{HADES:2020ver,NA49:2002pzu,NA49:2007stj,STAR:2017sal,ALICE:2013mez,ALICE:2019hno} on the collision energy dependence of charged pion multiplicity to parameterize its collision energy dependence from 2.4~GeV to 5.02~TeV.
We take a fit function from~\cite{ALICE:2013jfw} where it was used to parameterize the energy dependence of charged multiplicity.
The fit to the pion data yields
\begin{equation}
\label{eq:Npippim}
    \frac{dN_{\pi^+}}{dy} + \frac{dN_{\pi^-}}{dy} = a \, s_{\mathrm{NN}}^b \, \ln( s_{\mathrm{NN}}) - c.
\end{equation}
Here $s_{\mathrm{NN}}$ is in the units of GeV$^2$, and the parameter values are $a = 49.84903$, $b = 0.04110131$, and $c = 61.48846$.
Then, at each $\sqrt{s_{\mathrm{NN}}}$ we fix $V(\sqrt{s_{\mathrm{NN}}})$ to a value such that the thermal model reproduces $\frac{dN_{\pi^+}}{dy} + \frac{dN_{\pi^-}}{dy}$ from Eq.~\eqref{eq:Npippim}. 
We also check that total baryon number $dN_B/dy$ calculated at a given energy does not exceed the number of participants, $N_{\rm part} = 360$. If it does, the volume is rescaled down such that $dN_B/dy = N_{\rm part}$. This rescaling is only necessary at very low energies, $\sqrt{s_{\mathrm{NN}}} \lesssim 2.8$~GeV.

The effect of exact local conservation of strangeness becomes important for strange particles, such as hypernuclei, at low collision energies where the amount of the produced strangeness is small.
Here we incorporate this effect through the strangeness-canonical ensemble,
which enforces the exact conservation of net strangeness in a correlation volume $V_c$.
We take a correlation radius $R_c = 2.4$~fm~($V_c = \frac{4\pi}{3} R_c^3$), as inferred from the recent measurements of the $\phi/K^-$ ratio at $\sqrt{s_{\mathrm{NN}}} = 3$~GeV~\cite{STAR:2021hyx}.

We also use the thermal model to study the system-size dependence of light (anti-)(hyper-)nuclei ratios at LHC energies.
Canonical suppression effects drive this dependence.
Here we use the canonical statistical model of Ref.~\cite{Vovchenko:2018fiy}, with a constant temperature $T = 155$~MeV across all multiplicities, and the canonical correlation volume of $V_c = 1.6 \, dV/dy$ suggested by recent measurements of antiproton-antideuteron correlations~\cite{ALICE:2022amd}.

All our thermal model calculations are performed using the open-source Thermal-FIST package~\cite{Vovchenko:2019pjl}.
These calculations optionally include feed-down from the decays of excited nuclei, as described in~\cite{Vovchenko:2020dmv}.




 \subsection{Coalescence approach}

The coalescence approach to (hyper-)nuclei production assumes that these nuclei are produced after the kinetic freeze-out (last scattering or decay) of their constituents \cite{Aichelin:1991xy,Nagle:1994wj,Bleicher:1995dw,Nagle:1996vp,Puri:1996qv,Puri:1998te,Ko:2010zza,Botvina:2014lga,Zhu:2015voa,Botvina:2016wko,Sombun:2018yqh,Sun:2020uoj,Gaebel:2020wid,Zhao:2021dka,Kireyeu:2022qmv,Kittiratpattana:2020daw,Kittiratpattana:2022knq}. If the full phase space information on the nucleons and hyperons at this time is known, the probability of a pair or triplet of baryons forming a bound nucleus can be estimated from the coalescence formula \cite{Mattiello:1996gq}
\begin{widetext}
\begin{equation}
    \frac{\mathrm{d}N}{\mathrm{d}^3k} = g \int\mathrm{d}p_1^3\mathrm{d}p_2^3\mathrm{d}x_1^3\mathrm{d}x_2^3 f_A(p_1,x_1) f_B(p_2,x_2) \rho_{AB}(\Delta x, \Delta p) \delta(k - (p_1+p_2))
\end{equation}
\end{widetext}

where $f_A(p_1,x_1)$ and $f_B(p_2,x_2)$ are the phase space distributions of the constituent nucleons and $\rho_{AB}(\Delta x, \Delta p)$ is a phase space density determining the probability whether a bound state is formed. This density is often related to the wave function of the nucleus to be formed \cite{Mattiello:1996gq}. However, this may not be strictly true since the nuclei are not yet formed and the effective density $\rho_{AB}(\Delta x, \Delta p)$ may contain effects from final state interactions as well as the formation process. Note that, to avoid these issues, recent works have studied the effect of a dynamical cluster formation model \cite{Oliinychenko:2020znl} and it was found that the results are very similar to the coalescence framework when the two coalescence parameters $\Delta p$ and $\Delta r$ are adjusted appropriately.

In the present work we will employ a straightforward implementation of the coalescence procedure, named box-coalescence, based on freeze-out distributions obtained with the UrQMD transport model in cascade and hybrid mode. 

\begin{table}[b]
\centering
\begin{tabular}{|c|c|c|c|c|c|c|c|}
\hline
 & {NN} & $\Lambda \Lambda$ & $\Xi$N & NNN & $($NN$\Lambda)_a$ & $($NN$\Lambda)_b$ &  NN$\Xi$ \\
\hline
spin-isospin & 3/8 & 3/16 & 3/8 & 1/12 & 1/12 & 1/12 & 1/12\\
\hline
$\Delta r_{max}$ [fm] & 3.575 & 9.5 & 9.5 & 4.3 & 9.5 & 4.3 & 9.5 \\
\hline
$\Delta p_{max}$  [GeV] & 0.285 & 0.135 & 0.135 & 0.33 & 0.135 & 0.25 & 0.135 \\
\hline
\end{tabular}
\caption{Probabilities and parameters used in the UrQMD  phase-space coalescence.\label{tab:fac}}
\end{table}

The box-coalescence assumes a product of step-functions for $\rho_{AB}(\Delta x, \Delta p)$ where the probability for two nucleons $A$ and $B$ to form a bound state within a volume given by $\Delta p$ and $\Delta r$ is a constant and vanishes outside. This method has the advantage of being quick to compute and it was shown that it provides almost identical results to more complicated shapes of the probability density \cite{Nagle:1996vp}. 
The detailed procedure of the coalescence model was already described in \cite{Sombun:2018yqh,Bleicher:2019oog,Hillmann:2021zgj} for nuclei with two or three constituents. In short three steps are necessary:
\begin{enumerate}
\item Using the list of freeze-out coordinates from the UrQMD model, the relative coordinates of all nucleon (hyperon) pairs in their respective center of mass frame is calculated. If their relative distance $\Delta r=\left|\vec{r}_{n_1}-\vec{r}_{n_2}\right|<\Delta r_{max,nn}$ and momentum distance $\Delta p=|\vec{p}_{n_1}-\vec{p}_{n_2}|<\Delta p_{max,nn}$, a two nucleon state is potentially formed with the combined momenta $\vec{p}_{nn}=\vec{p}_{n_1}+\vec{p}_{n_2}$ at position ${\vec{r}_{nn}=(\vec{r}_{n_1}+\vec{r}_{n_2})/2}$.
\item  In a second step the local rest-frame of this two nucleon state and any other possible third nucleon or hyperon is calculated. If the conditions of their relative distance $\Delta r=|\vec{r}_{nn}-\vec{r}_{n_3}|<\Delta r_{max,nnn}$ and momentum distance $\Delta p=|\vec{p}_{nn}-\vec{p}_{n_3}|<\Delta p_{max,nnn}$ are fulfilled, a $A=3$ (hyper-)nucleus is formed with a probability given by the spin-isospin-coupling. The momentum of the three nucleon state is then $\vec{p}_{nnn}=\vec{p}_{nn}+\vec{p}_{n_3}$ and the position is $\vec{r}_{nnn}=\frac{1}{3} \left( \vec{r}_{n_1}+\vec{r}_{n_2}+\vec{r}_{n_3} \right)$.
\item If no third particle is found, a dibaryon can be formed with the appropriate probability given by the spin-isospin-coupling.
\end{enumerate}

The corresponding parameters are shown in table \ref{tab:fac}. Note that all parameters are energy independent. While the parameters for most light nuclei have been already fixed in previous studies \cite{Sombun:2018yqh,Bleicher:2019oog,Hillmann:2021zgj}, the parameters for the hypernuclei are not well constraint \cite{Schaffner-Bielich:1999zoi}. In the following we assumed two parameters sets $(a)$ and $(b)$. While $(a)$ is motivated by the large size of the wave function of the hypertriton and has a $\Delta r_{max} = 9.5$ fm, $(b)$ corresponds more to the triton size while the $\Delta p_{max}$ is adjusted to give the same hypertriton multiplicity at a beam energy of $\sqrt{s_{\mathrm{NN}}}= 20$ GeV.

\begin{figure}[t]
  \centering
  \includegraphics[width=0.5\textwidth]{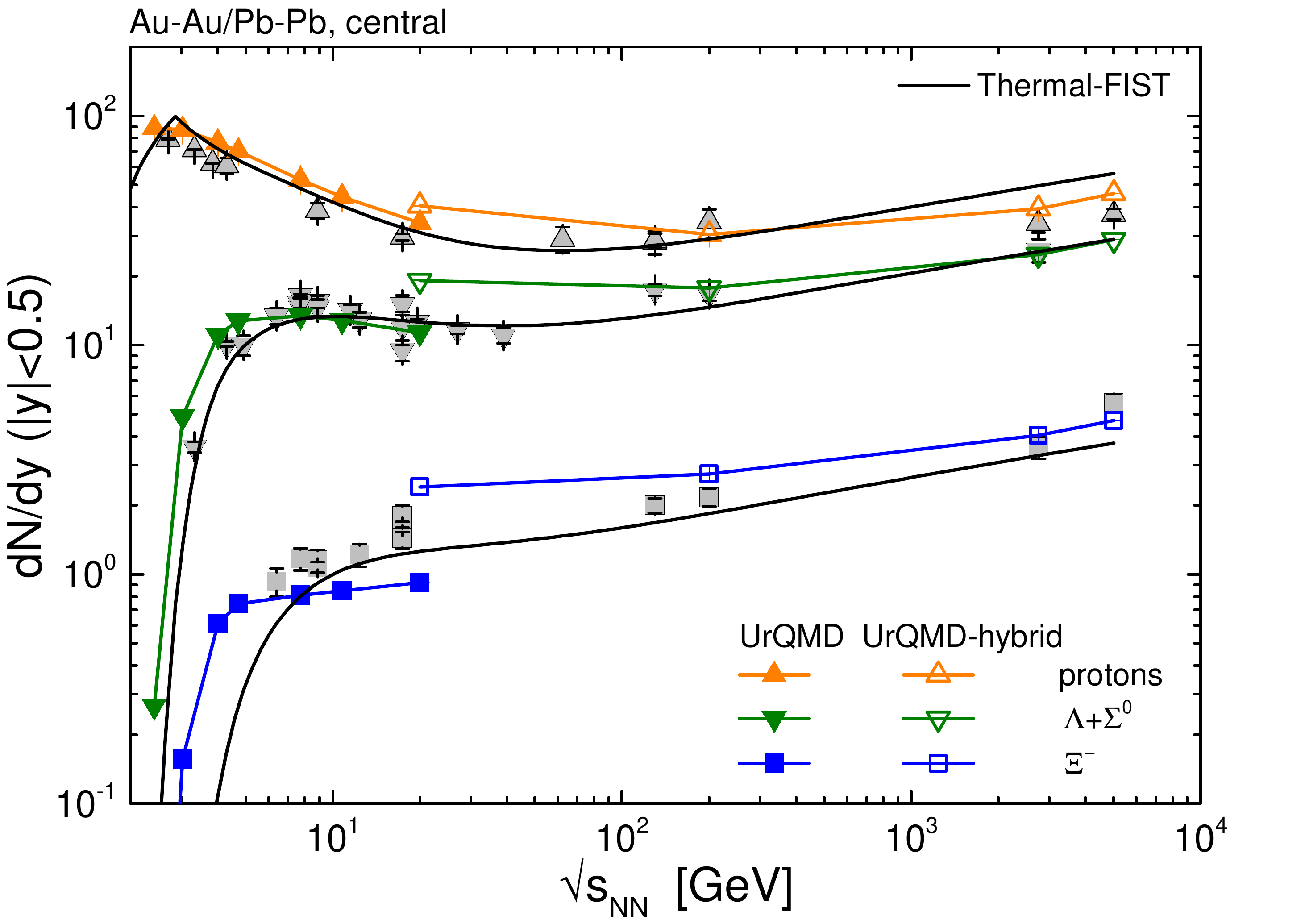}
  \caption{(Color online) Mid-rapidity yields of protons and hyperons in central collisions of heavy nuclei over a wide range of center of mass energies. The black lines correspond to thermal model results and the lines with symbols are results from the UrQMD transport model in cascade (full symbols) or hybrid (open symbols) mode. Experimental data is shown as grey symbols.
  }
  \label{fig:bar_yield}
\end{figure}

\section{Hadron multiplicities}

In both the Thermal-FIST and the UrQMD-coalescence approaches, the baryons, i.e. nucleon and hyperon multiplicities and their phase space distributions serve as input to the (hyper-)nuclei predictions. It is therefore necessary to provide a short overview of the capabilities and differences of these two approaches to describe the measured baryon multiplicities before turning to light nuclei. 
Figure \ref{fig:bar_yield} shows the measured and simulated mid-rapidity multiplicities for protons, $\Lambda$-hyperons as well as $\Xi$-hyperons from central collisions of heavy nuclei at beam energies ranging from $\sqrt{s_{\mathrm{NN}}}= 2$ GeV to $5$ TeV. The colored lines with symbols depict the UrQMD results, where the filled symbols represent the UrQMD-default and the open symbols the UrQMD-hybrid model. The thermal fit is shown as black lines. As expected the thermal fit gives a good description of the multiplicities, though there are several deviations especially at the highest beam energies. A distinct difference in these two models is the treatment of strangeness production. While the default UrQMD treats strangeness production microcanonically, which leads to a well known underestimation of multi-strange hadrons at higher beam energies, the hybrid model includes strangeness production from a Cooper-Frye procedure, assuming a grand canonical ensemble. In addition a slight overprediction of the proton number at midrapdity is observed in UrQMD for beam energies of  $\sqrt{s_{\mathrm{NN}}}> 5$ GeV which will likely be reflected in the final nuclei multiplicity.

\section{Results}

\begin{figure}[t]
  \centering
  \includegraphics[width=0.5\textwidth]{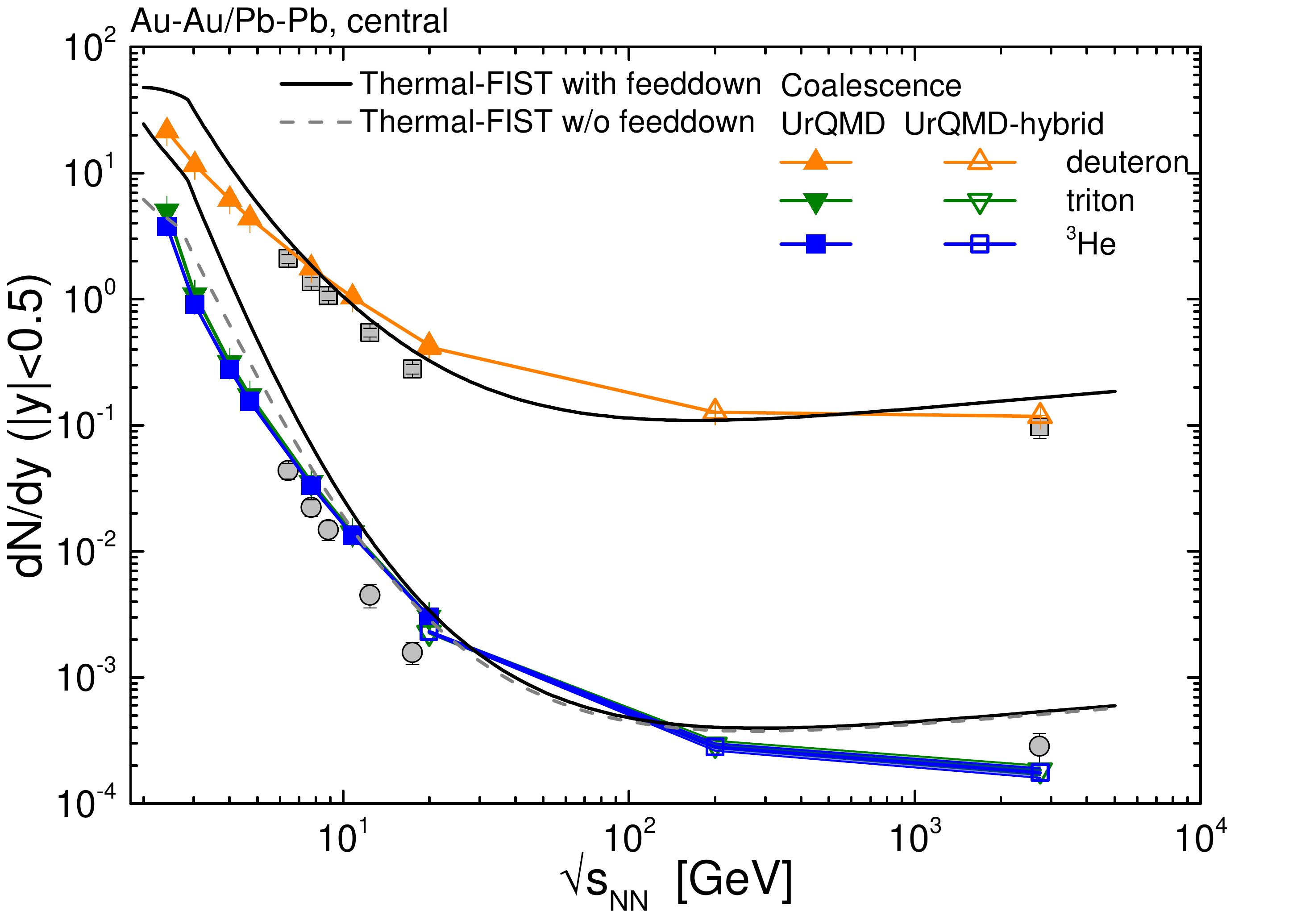}
  \caption{(Color online) Mid-rapidity yields of light nuclei for central collisions of heavy nuclei over a wide range of center of mass energies. The experimental data (grey symbols) is compared to thermal model results shown as black lines (including feed down from excited nuclei) and grey dashed line (without feed down). The lines with symbols denote results from the UrQMD model. For the thermal model only the  $^3$He and deuteron is shown. The difference between triton and $^3$He in UrQMD is only visible for the lowest beam energies due to the isospin imbalance in the projectile and target nuclei. 
  }
  \label{fig:nucl_yield}
\end{figure}

\begin{figure}[t]
  \centering
  \includegraphics[width=0.5\textwidth]{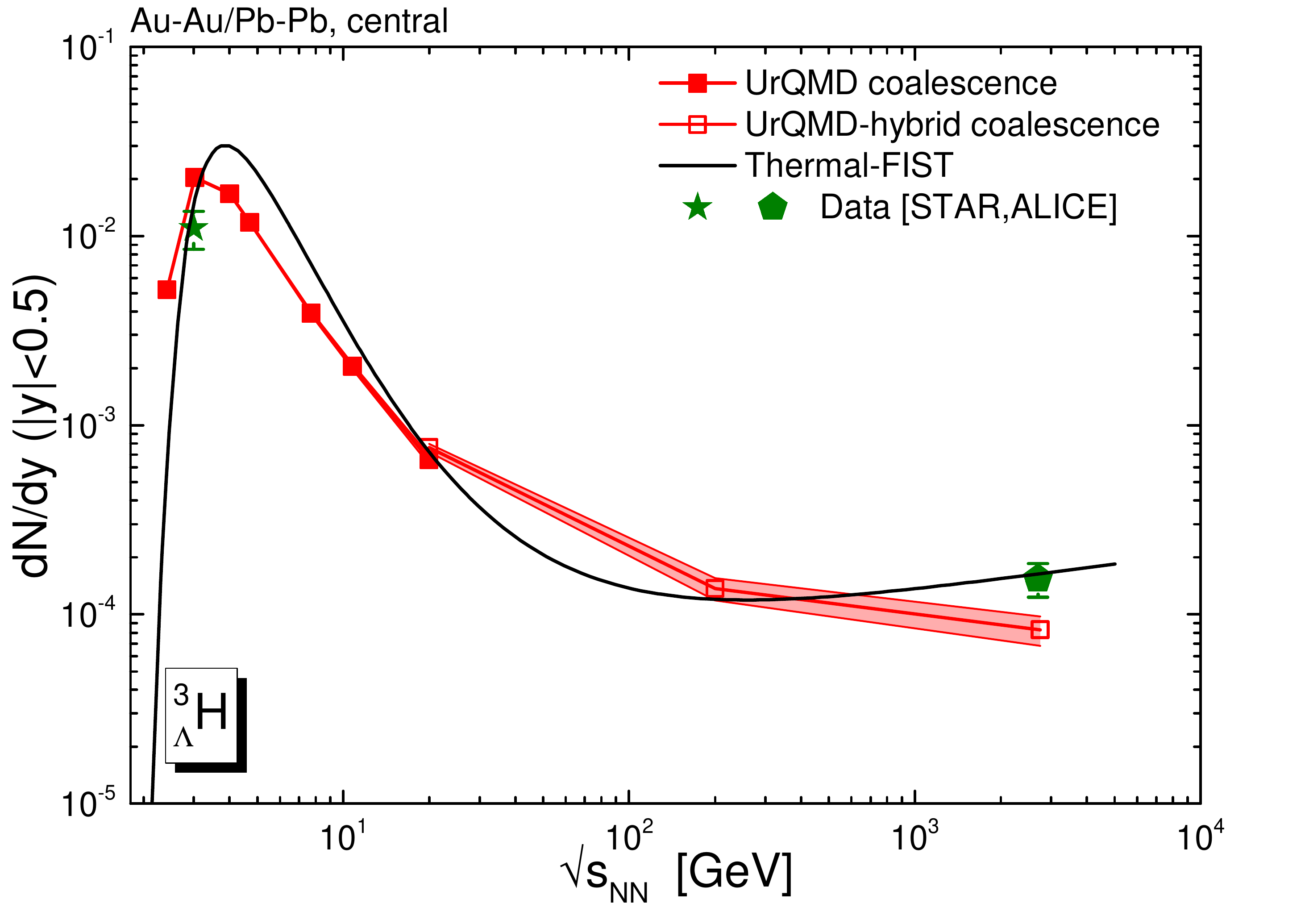}
  \caption{(Color online) Mid-rapidity yield of the hypertriton in central collisions of heavy nuclei as function of the center of mass collision energy. The thermal model prediction is compared with the coalescence results of the UrQMD model and available data from ALICE and STAR (preliminary).
  }
  \label{fig:hnucl_yield}
\end{figure}

\subsection{Multiplicities}

Before turning to the hypernuclear clusters, we first want to recapitulate the results on light nuclei production from the coalescence model. The beam energy dependence of ratios to protons of deuterons, tritons and $^3$He have been already published in \cite{Hillmann:2021zgj}. Figure \ref{fig:nucl_yield} shows the total mid-rapidity multiplicity, for central ($b<3.4$ fm) Au-Au or Pb-Pb collisions, of these three clusters as function of the beam energy. The coalescence results (colored symbols with lines) are compared with the Thermal-FIST results with (solid black line) and without (grey dashed line) the inclusion of feed-down from excited nuclei. Both theoretical results are compared with the available data (open symbols). Overall, both models seem to give a reasonably good description of the beam energy dependence. At the lower beam energy, the thermal model with feed-down predicts a larger multiplicity of light nuclei while at intermediate beam energies both models are almost identical. At the highest collision energy (the LHC) the thermal model predicts systematically more light nuclei than the coalescence description. This can have two reasons: 1. The annihilation of baryons and anti-baryons in the final hadronic stage has a stronger impact on the light clusters and therefore reduces their number significantly \cite{Steinheimer:2012rd,Sombun:2018yqh}. 2. The total mid-rapidity volume for cluster production in the UrQMD-hybrid simulation is smaller than the volume used in the thermal model at this collision energy. 

Next, we turn to the description of light hypernuclei. Figure \ref{fig:hnucl_yield} shows the result of the mid-rapidity multiplicity of hypertriton for central Au-Au or Pb-Pb collisions as function of collision energy. The red symbols with lines depict the coalescence results where we only show the multiplicity using parameter set I, which means a large $\Delta r$ and smaller $\Delta p$. Since there are only two data points available for the multiplicity it is not surprising that the coalescence model gives a reasonable description of the data. Similarly, the thermal model works well for the two available data points. It is however noteworthy, that the coalescence models predicts a slightly smaller peak in the hypertriton yields, due to the underestimation of the Lambda multiplicity at those beam energies in the UrQMD cascade model. In addition, for the highest beam energies again, a suppression of the multiplicity in the coalescence model compared to the thermal model is observed.  

\begin{figure}[t]
  \centering
  \includegraphics[width=0.5\textwidth]{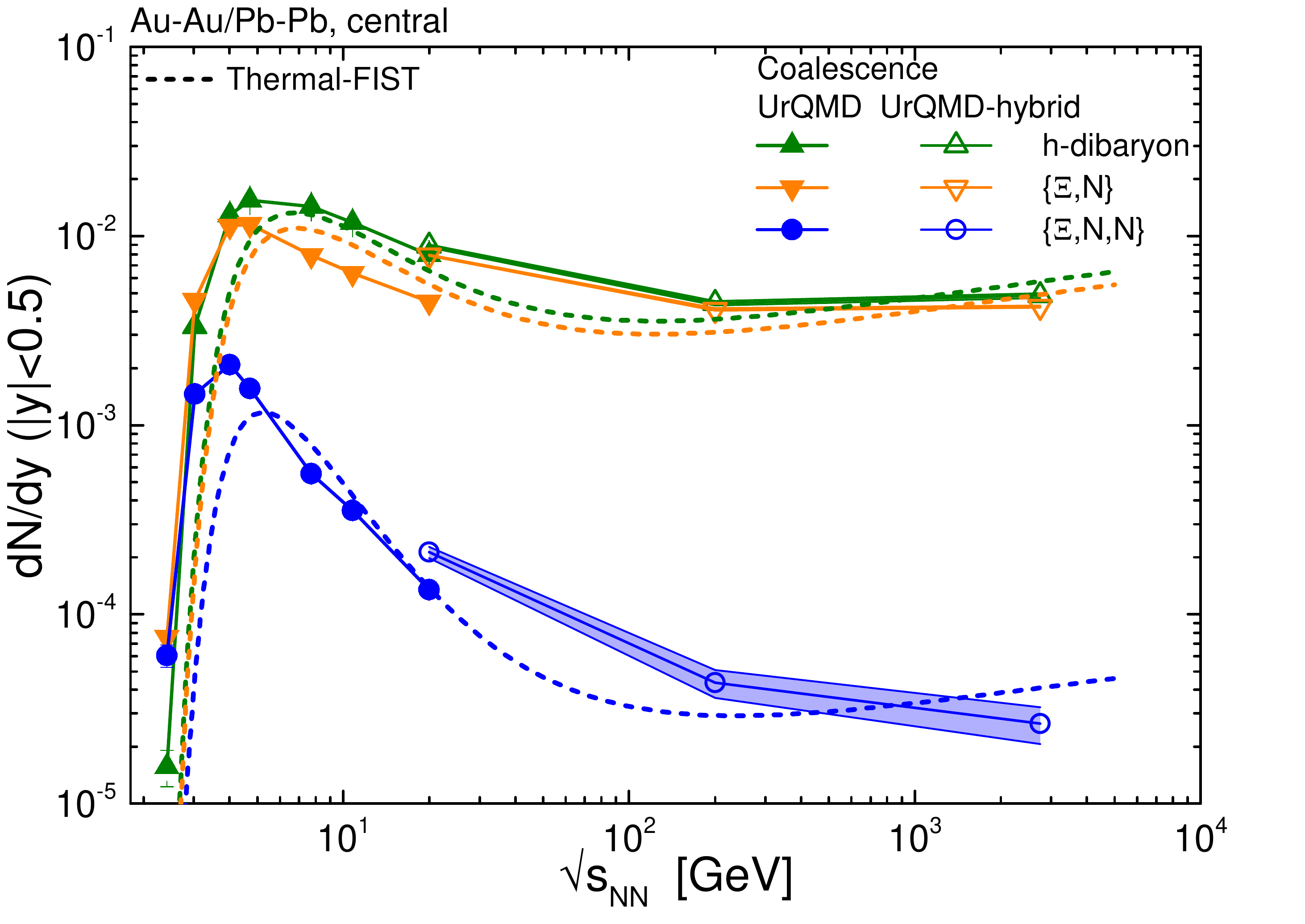}
  \caption{(Color online) Predicted mid-rapidity yields of different multi-strange light nuclei in central collisions of Au-Au/Pb-Pb as function of the center of mass collision energy. The thermal model predictions (dashed lines) are compared with the coalescence results of the UrQMD model. The multiplicity was estimated per isospin combination of the corresponding hypernuclear state. Due to the similar mass, the H-dibaryon and $\{ \Xi N \}$ have the same predicted multiplicity. 
  }
  \label{fig:2hnucl_yield}
\end{figure}

Finally, having fixed a reasonable parameter set for $\Delta r$ and $\Delta p$, for bound hyperons, we can use these parameters to predict other hypothetical but yet unconfirmed small hypernuclei. As such we will predict the multiplicity of the H-dibaryon ($\{ \Lambda, \Lambda \}$) as well as two possible bound states of the $\Xi$, the $\{ \Xi, N \}$ and $\{\Xi,N,N \}$. We show the multiplicity for the sum of all isospin combination in Fig.~\ref{fig:2hnucl_yield}. Again, the coalescence predictions (colored symbols with lines) are compared with the thermal model (dashed lines), both predict very similar multiplicities over a broad range of energies. Only for the lowest beam energies differences are observed due to the slightly different production of $\Lambda$ and $\Xi$ in the UrQMD model as compared to the thermal model. Also, at the LHC lower multiplicities are observed for the coalescence model than for the thermal model, similar to the other light nuclei. The fact that the $\{ \Lambda, \Lambda \}$ and  $\{ \Xi, N \}$ show almost identical multiplicities comes from their similar total mass, in the thermal model, and the relative multiplicities in the UrQMD model. Note, that $\{ \Xi, N \}$ includes a higher number of possible isospin combinations and therefore would, in total, have a higher multiplicity than an H-dibaryon. In general, we can conclude that, if they exist, the above predicted states would be frequently produced even at the LHC. The current best estimate for an upper limit on the H-dibaryon ($2\cdot 10^{-4}$ at 99\% CL \cite{ALICE:2015udw}), if it decays due to the weak interaction, is about one order of magnitude below our prediction which would rule out its existence. 

\begin{figure}[t]
  \centering
  \includegraphics[width=0.5\textwidth]{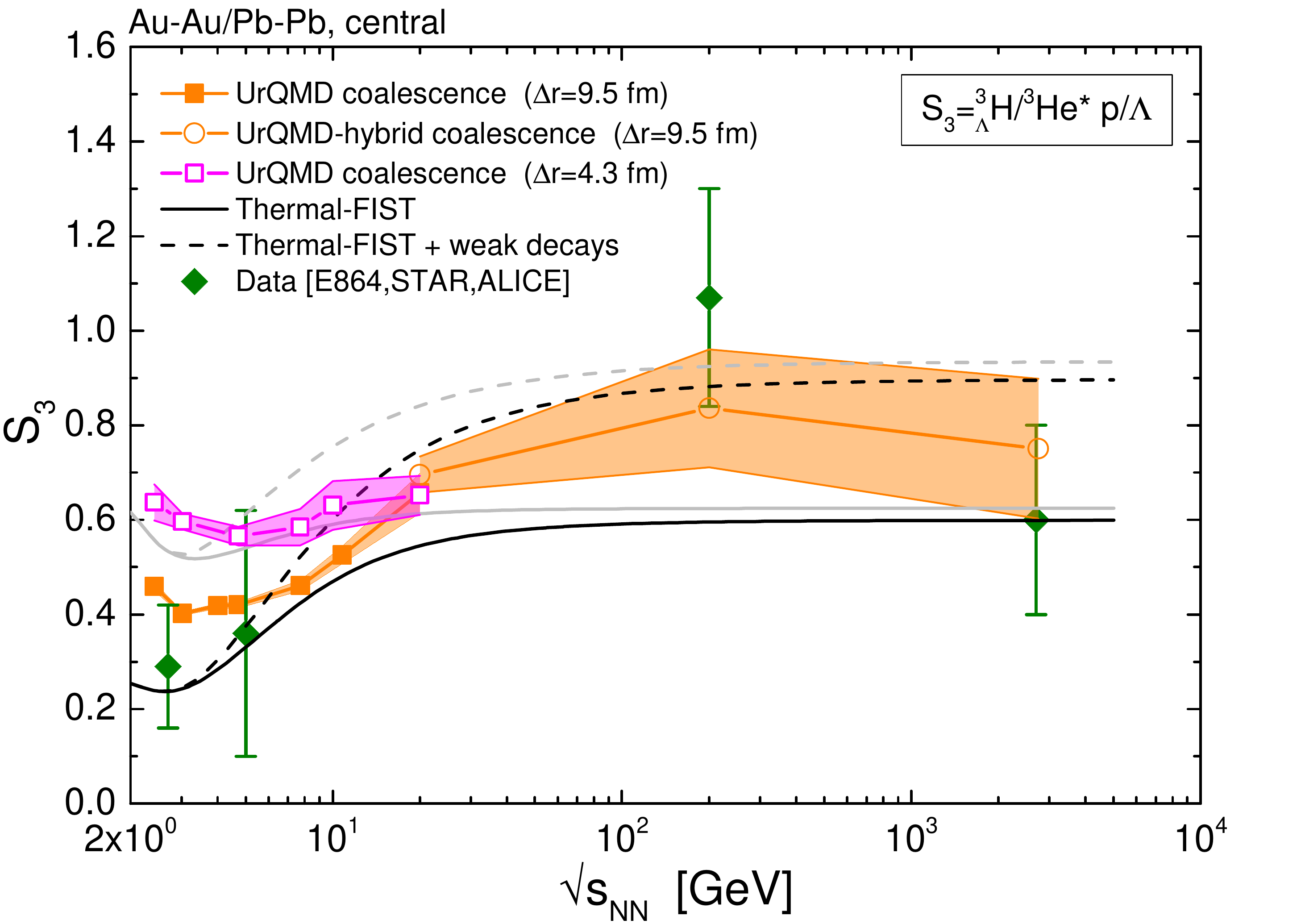}
  \caption{(Color online) Energy dependence of the double ratio $S_3={}_{\Lambda}^{3}\mathrm{H} / ^{3}\mathrm{He} \cdot p/(\Lambda+\Sigma^0)$. Several scenarios are compared. The black and grey lines correspond to thermal model estimates where the weak decay feed down from the $\Lambda$ to the proton is either taken into account (solid line) or not (dashed lines). If feed down from excited nuclei is included (black lines), the double ratio is reduced compared to a scenario where excited nuclear states are omitted (grey lines). The UrQMD+coalescence ($\Delta r = 9.5$) results are shown as orange line with full squares, UrQMD/hybrid+coalescence ($\Delta r = 9.5$) results are shown as orange line with circles. The results of UrQMD+coalescence ($\Delta r = 4.3$) are shown as magenta line with open squares. Data are shown by the green symbols.
  }
  \label{fig:s3}
\end{figure}

\subsection{Effects of the source size}

Until now it was simply assumed that the parameter $\Delta r$ which enters the coalescence prescription can be directly related to the size of the hypertriton wave function and thus $\Delta r = 9.5$ fm was chosen. However, this interpretation is not necessarily unique since in the coalescence the nuclei are 'created' directly at their point of last scattering, a point in space and time when they can be hardly be treated as an isolated system. This means also an interpretation of $\Delta r$ and $\Delta p$ as region of homogeneity or emission source, as in the pion HBT formalism, is possible.
To study the effects on how a change in this source size may effect the production probability we will modify the coalescence parameters. In particular we will study two scenarios: \textit{Set I} where $\Delta r = 9.5$ fm as suggested by the wave function size and \textit{Set II} where $\Delta r = 4.3$ fm as for the triton. The momentum distance is then adjusted to yield the same hypertriton multiplicity in central collisions at $\sqrt{s_{\mathrm{NN}}}= 20$ GeV. To understand why such a modification can change the yield of hypernuclei, even in a picture where the wave function is not involved, one can consider a simple example: Since the momentum vector at the emission time of nuclei usually points outwards, i.e. it is correlated with the position vector, even for systems with small flow. If the freeze-out hypersurface is very large, as compared to the source volume of the (hyper-)nucleus, the position and momentum of emitted hadrons are then correlated. If the freeze-out surface is smaller and has a significant curvature, this correlation is reduced and only hadrons which are close in coordinate space are likely to have momenta in the same direction. In fact, one could even argue that the region of homogeneity resembles a Gaussian shape, as for the pion HBT, due to the thermal smearing on the hypersurface, mimicking what is usually used as 'wave function' of the nucleus.

\begin{figure}[t]
  \centering
  \includegraphics[width=0.5\textwidth]{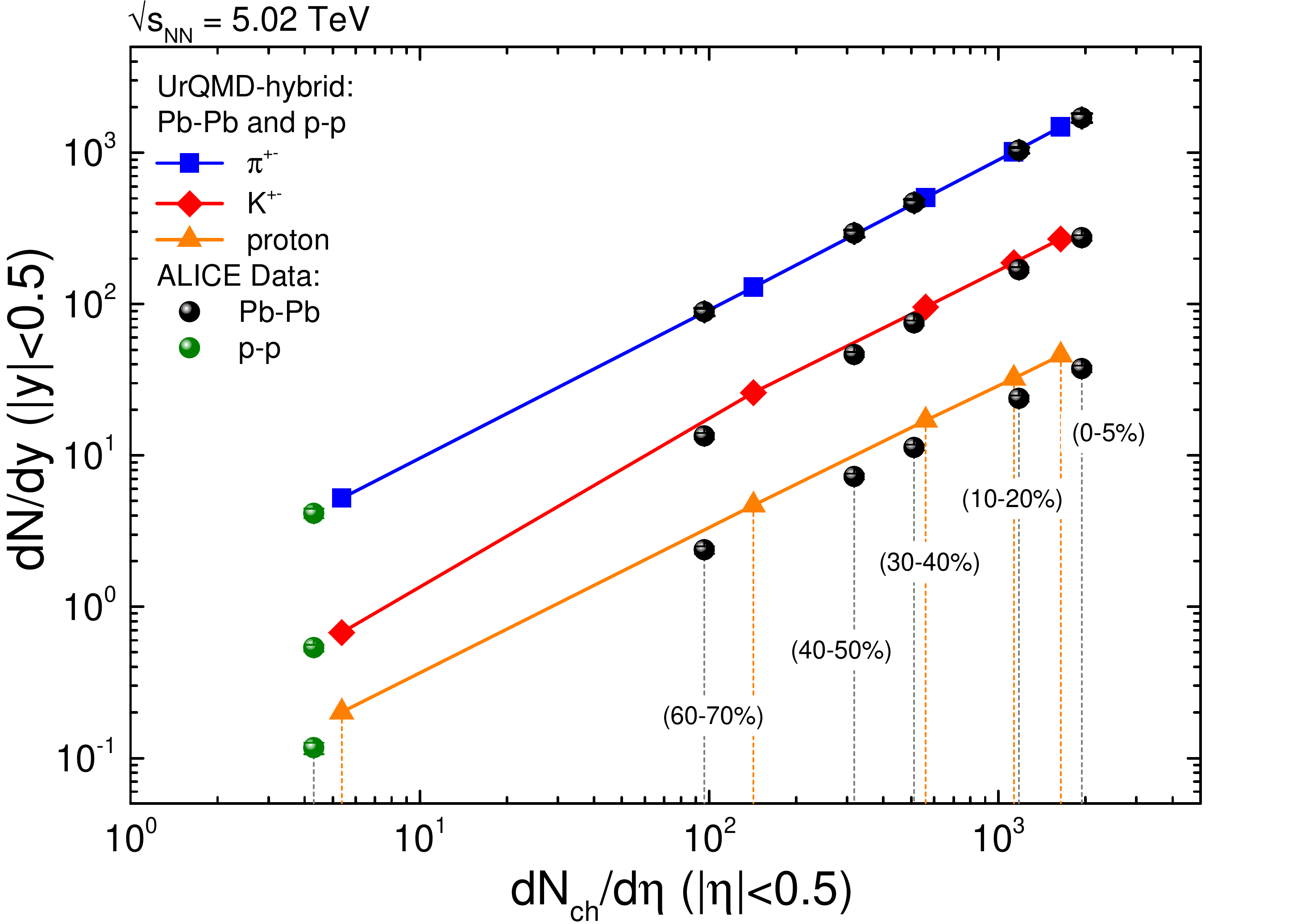}
  \caption{(Color online) Multiplicity dependence of different hadrons for Pb-Pb and p-p collisions at $\sqrt{s_{\mathrm{NN}}}=5.02$ TeV. The UrQMD-hybrid simulation results are compared to experimental data from the ALICE collaboration. 
  }
  \label{fig:had_cent}
\end{figure}

The ratio $S_3={}_{\Lambda}^3\mathrm{H}/^3\mathrm{He} \cdot p/\Lambda$ is very useful to study the differences in light nuclei and hypernuclei production, since it cancels out any effects from the different production of the hyperon involved. Figure \ref{fig:s3} presents the prediction of $S_3$ from the thermal model with (black lines) and without feed-down from excited nuclei (grey lines). The dashed lines are added to depict what would be expected if the proton number is not corrected for the weak decay feed-down from the hyperons. The coalescence results from UrQMD, using \textit{set I}, are depicted as orange symbols with error band. The results for \textit{set 2} are shown as magenta symbols with band.

Several observations can be made. The experimental data at different beam energies seem inconsistent, a problem which may be related to different feed-down corrections employed. Besides this the thermal model, including the excited nuclear states, gives a good description of most data, even though it overestimates nuclei production at the lowest beam energies. The coalescence model with UrQMD using \textit{set I} appears to give the best description of the data. Here, the drop of $S_3$ at lower beam energies is due to the large source size of the hypernucleus.
Due to the lack of high precision data it is difficult to draw any final conclusions from this comparison. However, there is another possibility to change the system size as compared to the emission volume by varying the centrality \cite{Sun:2018mqq,Vovchenko:2018fiy}.

\begin{figure}[t]
  \centering
  \includegraphics[width=0.5\textwidth]{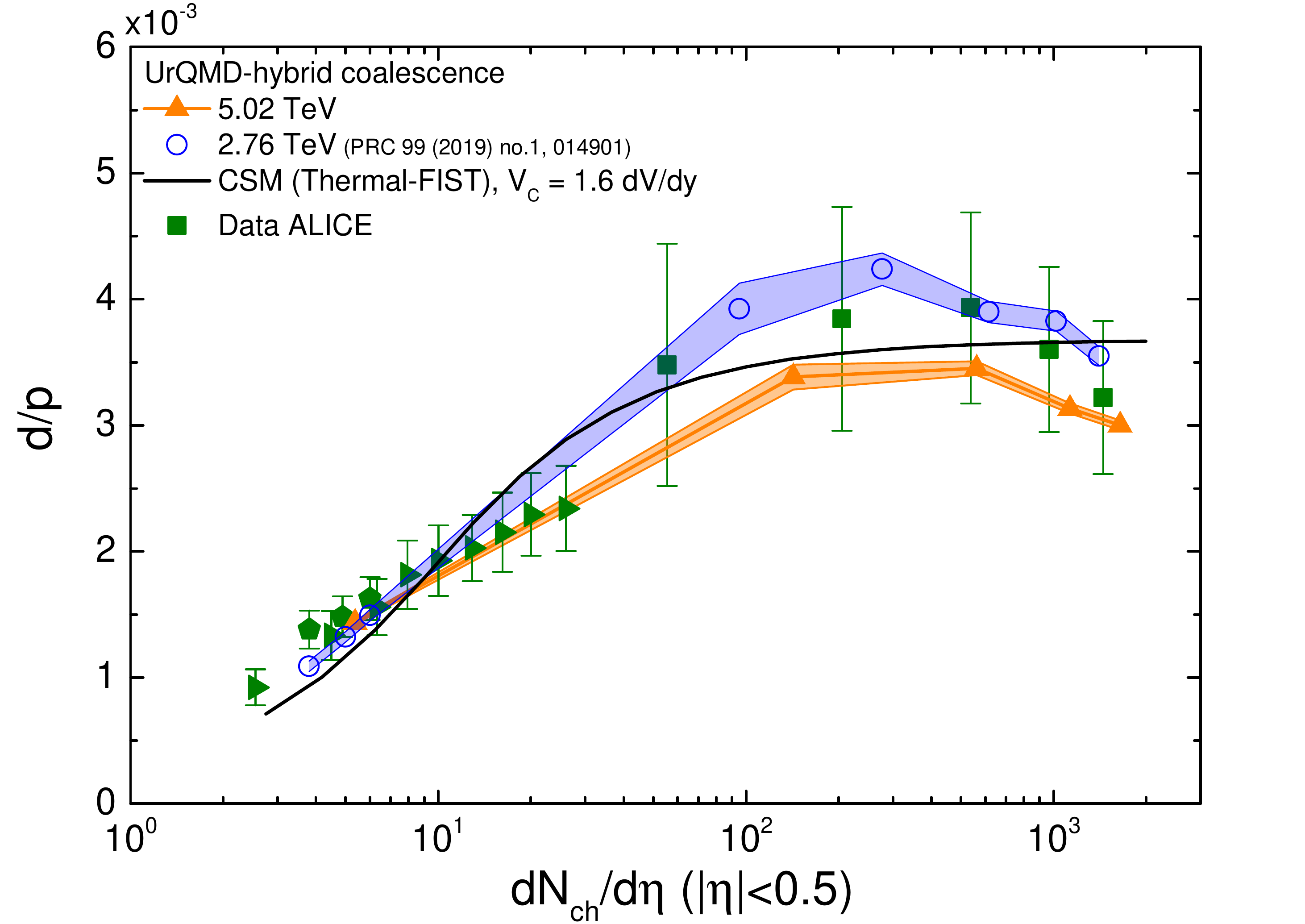}
  \caption{(Color online) Multiplicty dependence of the deuteron to proton ratio for $\sqrt{s_{\mathrm{NN}}}=5.02$ and $2.76$ TeV. The new UrQMD-hybrid simulations are shown as orange band and compared to previous UrQMD results \cite{Sombun:2018yqh} (blue band) as well as experimental data from the ALICE experiment. The prediction from the canonical thermal model fit is shown as solid black line.
  }
  \label{fig:deut_cent}
\end{figure}

\subsection{Centrality dependence}

Varying the centrality of a collision system is a useful method to change its volume without modifying the chemical composition. Especially, at the highest beam energies, the ALICE experiment has recently published a wealth of data on the centrality dependence on light nuclei. In the following we will compare results of (hyper-)nuclei production as function of the collision centrality from the coalescence model in the UrQMD-hybrid model with thermal model results that also incorporate canonical effects. To set the stage, Fig.~\ref{fig:had_cent} shows a compilation of hadron multiplicities as function of the number of charged particles in mid-rapidity. The experimental data (symbols) from Pb-Pb and p-p collisions is compared to UrQMD-hybrid model results of the same systems and the same centralities, defined by the impact parameter range obtained from a Glauber model \cite{ALICE:2013hur,Loizides:2016djv,Loizides:2017ack,dEnterria:2020dwq}. While the overall trend is nicely reproduced we already see that there is a general shift in the number of charged particles for a given centrality range. In addition it is observed that the UrQMD-hybrid model predicts a slightly too large proton multiplicity even though baryon-antibaryon annihilations are included in the hadronic afterburner of the simulation. Taking these caveats into consideration, we will now study the centrality dependence of (hyper-)nuclei production within the UrQMD+coalescence framework.

First, Fig.~\ref{fig:deut_cent} shows the deuteron to proton ratio as function of charged-particle multiplicity at midrapidity for Pb-Pb collisions at $\sqrt{s_{\mathrm{NN}}}=5.02$ and $2.76$ TeV (UrQMD data taken from a previous publication \cite{Sombun:2018yqh}) compared to data from ALICE and the canonical statistical model~(CSM) fit. The thermal model uses a chemical freeze-out temperature, which is fixed for the most central collisions, of $T=155$ MeV. The centrality dependence is then uniquely determined by the canonical freeze-out rapidity-volume which was chosen as $V_c=1.6 dV/dy$ \cite{ALICE:2022amd}. The canonical effects lead to a significant reduction of the d/p ratio for peripheral collisions which is also observed in the hybrid model simulation which uses a similar size initial rapidity volume as the canonical fit. In addition to the canonical suppression, the hybrid simulation also includes final state $B\overline B$ annihilations which lead to an additional suppression of the d/p ratio for the most central collisions, an effect which is also consistent with the shown ALICE data. 

\begin{figure}[t]
  \centering
  \includegraphics[width=0.5\textwidth]{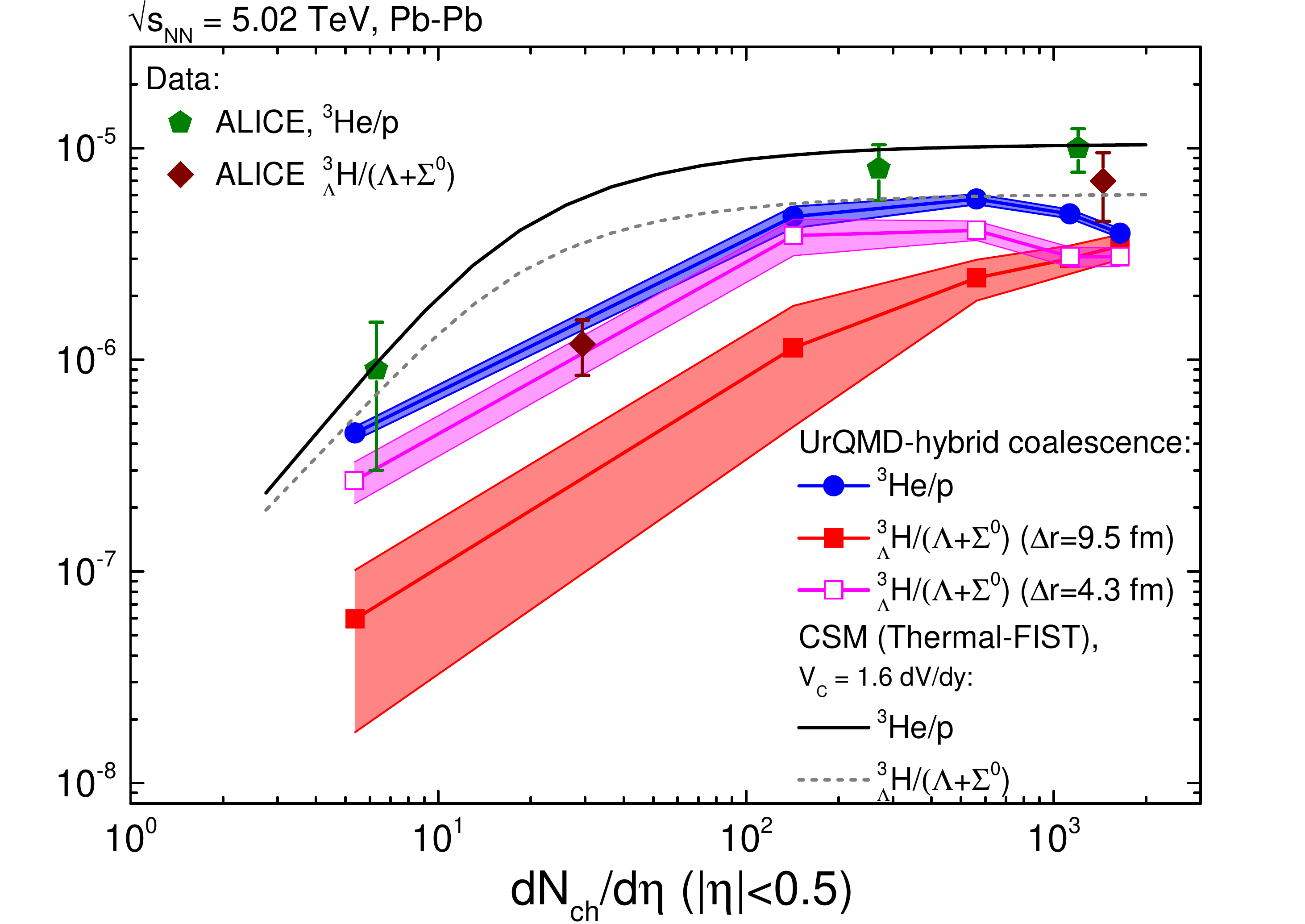}
  \caption{(Color online) Multiplicity dependence of the $^{3}$He to proton and $^{3}_{\Lambda}$H to $\Lambda$ ratio for Pb-Pb collisions at $\sqrt{s_{\mathrm{NN}}}=5.02$ TeV using the UrQMD-hybrid model. The $^{3}$He to proton ratio shows an expected maximum due to the increased annihilation of baryons in central collisions. In the case of the $^{3}_{\Lambda}$H to $\Lambda$ ratio we show two different scenarios corresponding to two different choices of the $\Delta r$ parameter.  The predictions from the canonical thermal model fits are shown as solid black line ($^3$He/p) and dashed grey line ($^{3}_{\Lambda}$H/($\Lambda + \Sigma^0$)).
  }
  \label{fig:rat_cent}
\end{figure}

Similarly, one would expect to observe such a suppression, due to the annihilations, also in the charged-particle multiplicity dependence of the $^3$He/p ratio. Figure \ref{fig:rat_cent} shows this dependence clearly as blue line together with the predictions from the CSM thermal results. Again the ratio is strongly suppressed in peripheral collisions due to canonical effects.
In this case the ALICE data do not show a suppression for the most central collisions, however, the few data points available may be not sufficient for any conclusions. When turning to the ratio of hypertriton to $\Lambda$, shown as red and magenta lines in Fig.~\ref{fig:rat_cent}, an interesting trend can be observed. When the coalescence radius for the hypertriton is large, $\Delta r = 9.5$ fm, the ratio rapidly drops towards peripheral collisions and does not show any constant or even peak behavior as for the $^3$He. When the coalescence radius is chosen equal to that of the $^3$He, a qualitatively similar behavior to the $^3$He can be observed. The rapid drop of the $^3_{\Lambda}$H/$\Lambda$ ratio was discussed as a direct result of the size of the coalescence distance in an analytical coalescence approach in \cite{Sun:2018mqq}. In the present work this effect can be quantitatively confirmed. Unfortunately, the currently available data is not sufficient to exclude any parameter set as the central coalescence yield can always be scaled up or down by another choice of the relative momentum. Since also the canonical effects lead to a decrease for peripheral collisions clear statements on the possible rapid drop or even peak like behavior of the ratio require much more precise data for several rapidity selections.

\begin{figure}[t]
  \centering
  \includegraphics[width=0.5\textwidth]{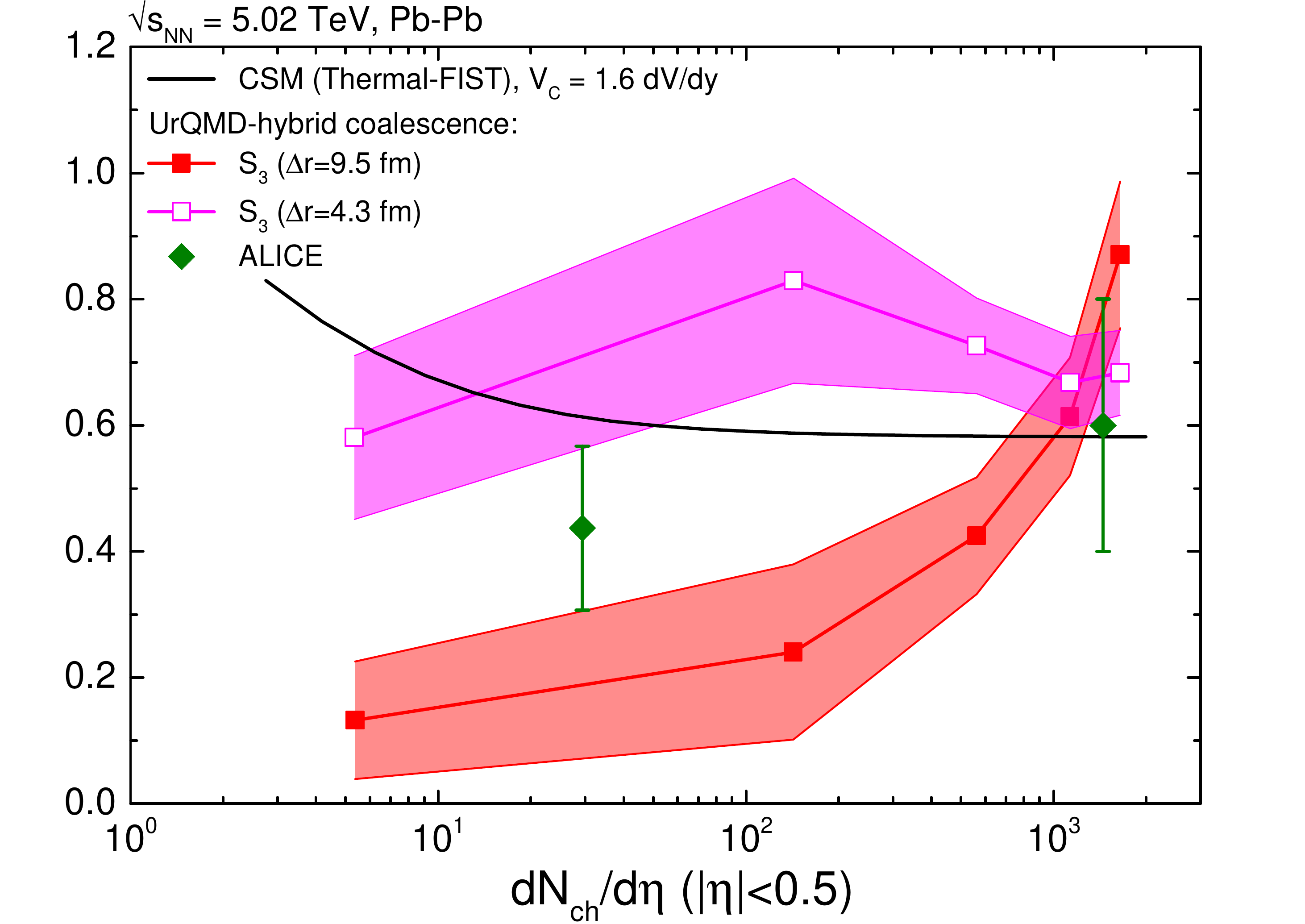}
  \caption{(Color online) Multiplicity dependence of the $S_3$ double ratio for Pb-Pb collisions at $\sqrt{s_{\mathrm{NN}}}=5.02$ TeV using the UrQMD-hybrid model. We compare two different scenarios corresponding to two different choices of the $\Delta r$ parameter for the hypertriton.  The predictions from the canonical thermal model fits are shown as solid black line.
  }
  \label{fig:S3_cent}
\end{figure}

To avoid the complication by the canonical effects it was suggested to use the double ratio $S_3$ instead of single ratios. In case of the double ratio, the canonical effects are mostly canceled and even lead to a small increase of $S_3$ for peripheral collisions, as seen in the CSM results in Fig.~\ref{fig:S3_cent}. In case of the same coalescence radius for $^3$He and hypertriton, $S_3$ is also essentially independent of centrality but shows a strong decrease if the radius is larger for the hypertriton. This constitutes a clear qualitative signal which can indicate whether the source volume for the hypernuclei is smaller or larger than for the normal nuclei. Unfortunately the ALICE data does not allow any clear distinction due to few available data with large error bars (green symbols).

\begin{figure}[t]
  \centering
  \includegraphics[width=0.5\textwidth]{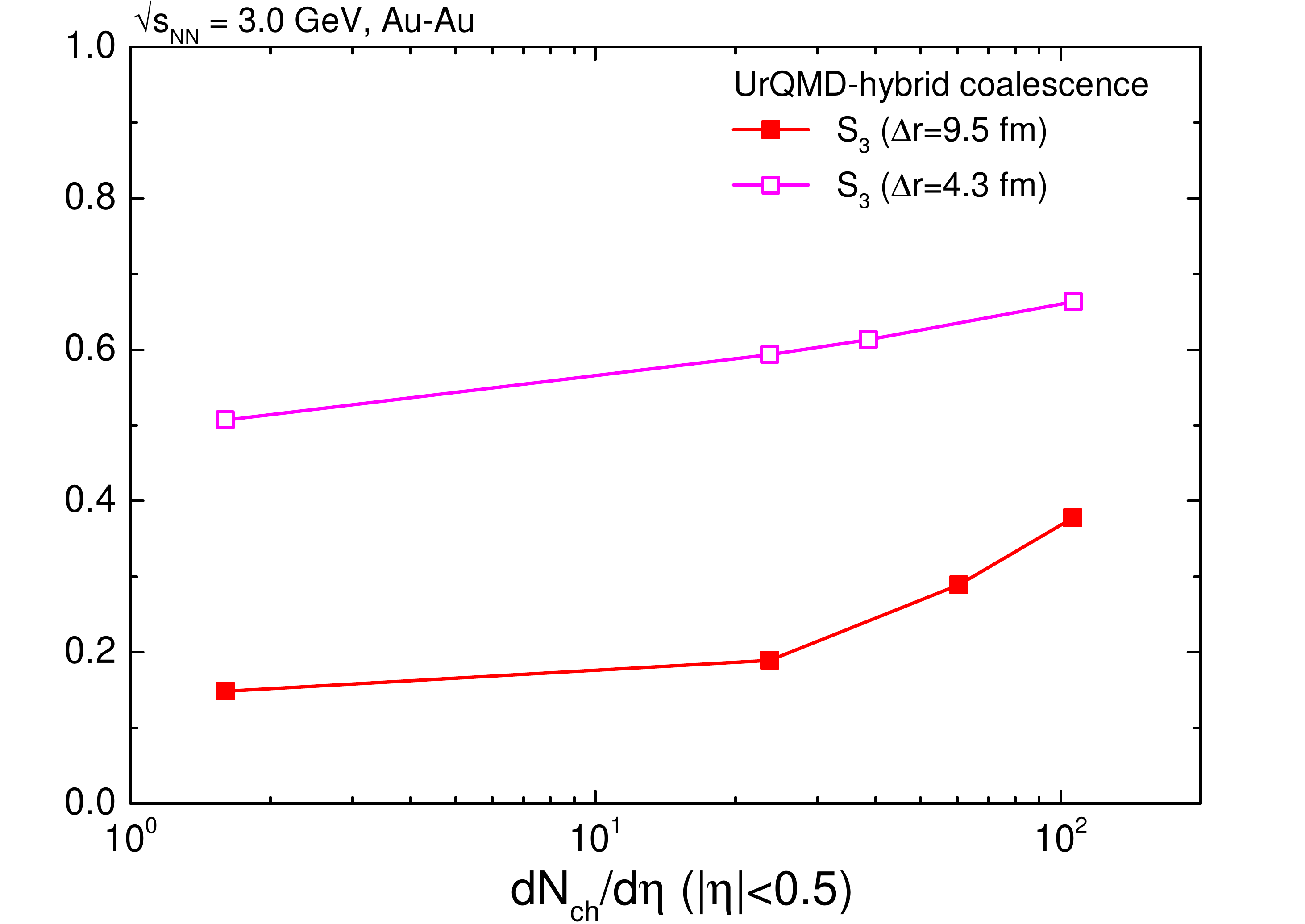}
  \caption{(Color online) Multiplicity dependence of the $S_3$ double ratio for Pb-Pb collisions at $\sqrt{s_{\mathrm{NN}}}=3.0$ GeV using the UrQMD-hybrid model. We compare two different scenarios corresponding to two different choices of the $\Delta r$ parameter for the hypertriton.
  }
  \label{fig:S3_cent_3gev}
\end{figure}

It should be noted however that we do not necessarily have to rely on the measurement at LHC energies where nuclei are very rare probes. Figure \ref{fig:S3_cent_3gev} shows the double ratio $S_3$ as function of centrality for Au-Au collisions at a much lower beam energy of $\sqrt{s_{\mathrm{NN}}}=3.0$ GeV which are being investigated by the STAR experiment. Again, a different source size or coalescence radius leads to a significant reduction of $S_3$ for peripheral collisions as compared to the central value. What makes the situation a bit more complex here is the fact that also the value in central collisions is already modified by the source radius observed at this low beam energy (see also Fig. \ref{fig:s3}).

\subsection{Elliptic flow scaling}

\begin{figure}[t]
  \centering
  \includegraphics[width=0.5\textwidth]{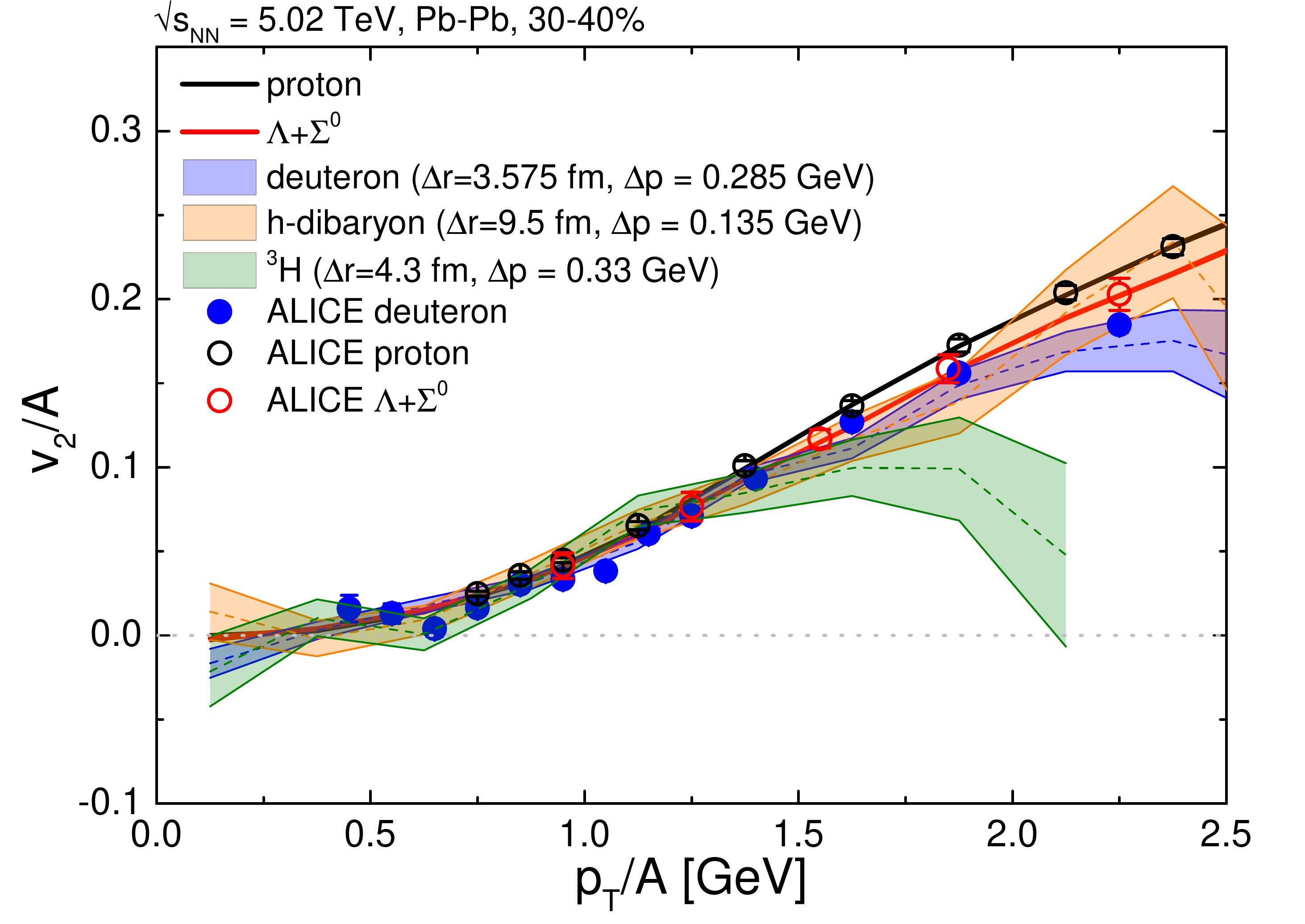}
  \caption{(Color online) Scaled elliptic flow of protons (solid black line), $\Lambda$'s (red solid line), deuterons (blue band), H-dibaryon (orange band) and triton (green band) for mid-central collisions of Pb-Pb at $\sqrt{s_{\mathrm{NN}}}=5.02$ TeV. Within errors elliptic flow scales for all light nuclei, independent of their size parameter $\Delta r$. For the deuteron and triton an increasing deviation from the mass scaling is observed for larger transverse momentum, which may also be observed in the ALICE data (open and closed colored symbols).
  }
  \label{fig:v2_A}
\end{figure}

The scaling of flow with the mass number of light nuclei is a direct consequence of the coalescence approach and was demonstrated in previous publications. What is not clear is whether this scaling is true exactly or if there are small deviations due to the relative momenta of the nucleons and if the scaling is still obtained for large coalescence parameters which allow the constituents to come from different parts of the fireball. To investigate this, the elliptic flow of protons, hyperons, deuterons as well as H-dibaryons and tritons is calculated within the UrQMD hybrid model for the top LHC beam energy. A centrality class of 30-40$\%$ central collisions was selected and the elliptic flow was calculated with respect to the reaction plane of the simulation according to 
\begin{equation}
    v_2 = \left\langle\frac{p_x^2 - p_y^2}{p_x^2 + p_y^2} \right\rangle ~,
\end{equation}
where the average runs over all particles in the mid-rapidity, $|y|<0.5$, region of the collision. The resulting scaled elliptic flow $v_2/A$ is shown as function of the scaled transverse momentum $p_T/A$ in Fig.~\ref{fig:v2_A}. The UrQMD hybrid model results are compared to experimental data on protons and hyperons as well as deuterons from the ALICE collaboration. It is clear that the elliptic flow scales with the mass number for essentially all light (hyper-)nuclei. Small deviations are observed between the protons and hyperons, which appear in the simulations as well as the data. In addition the scaling seems to break for higher momenta and this breaking appears earlier in terms of the scaled momentum as seen for the $A=3$ nucleus triton. We do not observe any significant difference in the $A=2$ nuclei deuteron and the H-dibaryon, even though they are produced with widely different coalescence parameters. This indicates that the flow scaling is not very sensitive to the coalescence parameters.

\section{Conclusion}

A comprehensive collection of model predictions for light hypernuclei production in relativistic heavy ion collisions was presented. Both the coalescence approach, applied after the kinetic freeze-out in the microscopic UrQMD model, as well as the thermal model (Thermal-FIST) were able to describe hypertriton production yields as measured by the STAR experiment. Using the parameters for these data the expected multiplicities for hypothetical light multistrange hypernuclei were presented. From this prediction the existence of the H-dibaryon ($\Lambda\Lambda$) seems ruled out by ALICE data \cite{ALICE:2015udw,ALICE:2019eol}. Furthermore we discussed the role of the coalescence parameter as well as canonical effects in the system size dependence and elliptic flow of light nuclei and hypertriton production. It was found that, while the ratios of hypertriton to $\Lambda$ are affected by both the source radius $\Delta r$ of the coalescence procedure as well as canonical effects. 
On the other hand, the double ratio $S_3$ is almost independent of canonical effects, which is in contrast to coalescence. 
While the elliptic flow was shown to be unaffected by the source size of the nuclei, an almost perfect mass scaling of the elliptic flow was observed which breaks down for large transverse momenta. It was found that, both the beam energy dependence and centrality dependence of $S_3$ can be used to constrain the hypertriton source radius. To do so the currently available data are not yet sufficient. More detailed studies about scaling of various nuclei at $\sqrt{s_\mathrm{NN}}=3$ GeV will be conducted in the future.

\begin{acknowledgments}
M.B. acknowledges support by the EU-STRONG 2020 network. B.D. acknowledges the support from Bundesministerium für Bildung und Forschung through ErUM-FSP T01 (F\"orderkennzeichen 05P21RFCA1).
V.V. was supported through the U.S. Department of Energy, 
Office of Science, Office of Nuclear Physics, under contract number 
DE-FG02-00ER41132. 
The computational resources for this project were provided by the Center for Scientific Computing of the GU Frankfurt and the Goethe-HLR.
\end{acknowledgments}



\bibliography{refs}

\end{document}